\pdfoutput=1
\documentclass[%
11pt, preprint, linenumbers, notitlepage,
 amsmath,amssymb,
 aps, physrev,
]{revtex4-2}

\usepackage{graphicx}
\usepackage{subcaption}
\usepackage{dcolumn}
\usepackage{bm}

\begin{document}

\abovedisplayskip=4pt
\abovedisplayshortskip=2pt
\belowdisplayskip=4pt
\belowdisplayshortskip=2pt
\abovecaptionskip=2pt
\belowcaptionskip=2pt
\nolinenumbers

\title{%
    {\Large Estimating detector systematic uncertainties for the T2K far detector} \\
    {\small Contribution to the 25th International Workshop on Neutrinos from Accelerators}
}%

\author{Michael Reh -- for the T2K Collaboration}
\email{michael.reh@colorado.edu}
\affiliation{Department of Physics, University of Colorado Boulder}



\begin{abstract}
    Tokai to Kamioka (T2K) is a long-baseline neutrino oscillation experiment that measures oscillation parameters related to both $\nu_\mu(\bar{\nu}_\mu)$ disappearance and $\nu_e(\bar{\nu}_e)$ appearance in a $\nu_\mu(\bar{\nu}_\mu)$ beam. T2K uses Super-Kamiokande (SK) as its far detector, and SK detector systematic errors are currently among the leading sources of systematic uncertainty in the T2K oscillation analysis. Therefore, accurate understanding of detector mis-modelling and event reconstruction uncertainties in SK is crucial to the extraction of the neutrino oscillation parameters. The detector error estimation procedure quantifies uncertainties by fitting SK atmospheric MC to data in a Markov Chain Monte Carlo (MCMC) framework. Uncertainties are separated based on true event topology above Cherenkov threshold in the SK atmospheric MC in order to capture how the uncertainties should affect the different signal and background samples when propagated to the T2K beam oscillation analysis. The procedure has been upgraded for the upcoming analysis cycle to include systematics targeting newly added $\nu_eCC1\pi^\pm$ and $NC\pi^0$-enhanced analysis samples, among other changes, culminating in a 540 dimensional MCMC fit between atmospheric MC and data in SK. The far detector analysis chain and its integration into the broader T2K oscillation analysis will be discussed in these proceedings.
\end{abstract}


\maketitle
\newpage


\section{\label{sec:t2k}The T2K Experiment}

The T2K experiment \cite{t2kexperiment} is a long-baseline neutrino oscillation experiment based in Japan that observes $\nu_\mu(\bar{\nu}_\mu)$ disappearance and $\nu_e(\bar{\nu}_e)$ appearance in a $\nu_\mu(\bar{\nu}_\mu)$ beam. T2K utilizes the J-PARC accelerator in Tokai, Japan to produce a narrow-band $\nu_\mu$ beam peaked at 0.6~GeV. This neutrino beam is sampled by a suite of near detectors that tune the predicted beam flux and interaction cross-section models before traveling 295~km west towards the far detector, Super-Kamiokande (SK) \cite{skdetector}.

\begin{figure*}[b!]
    \centering
    \begin{subfigure}[b]{0.45\textwidth}
        \centering
        \includegraphics[height=50mm]{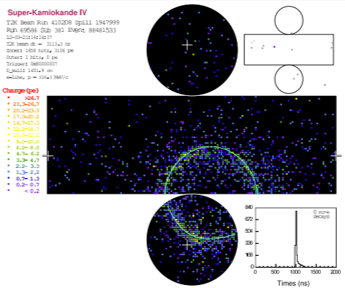}
        \caption{$\nu_e$ candidate event in SK, characterized by a ``fuzzy'' outline.}
        \label{subfig:nue}
    \end{subfigure}
    ~
    \begin{subfigure}[b]{0.45\textwidth}
        \centering
        \includegraphics[height=50mm]{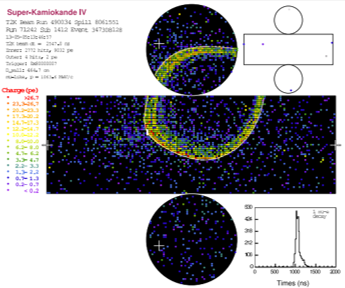}
        \caption{$\nu_\mu$ candidate event in SK, characterized by a ``sharp'' outline.}
        \label{subfig:numu}
    \end{subfigure}
    \caption{Cherenkov rings in Super-Kamiokande from the T2K run 1-4 data set.}
    \label{fig:example_rings}
\end{figure*}

SK is a 50 kT water Cherenkov detector located beneath Mount Ikeno in western Japan. As it relates to T2K, the 0.6~GeV neutrino beam peak energy combined with the 295~km baseline between J-PARC and SK makes it maximally likely for $\nu_\mu(\bar{\nu}_\mu)$ to disappear and for $\nu_e(\bar{\nu}_e)$ to appear at SK. As T2K extracts the neutrino oscillation parameters by observing the oscillated $\nu_\mu(\bar{\nu}_\mu)$ and $\nu_e(\bar{\nu}_e)$ spectra, it is extremely important that muons and electrons (and other particles, for that matter) can be accurately distinguished in the detector. Generally, electrons leave characteristic ``fuzzy'' rings in SK as in Figure~\ref{subfig:nue}, while muons leave ``sharp'' rings as in Figure~\ref{subfig:numu}, which the T2K event reconstruction can reliably discern. Even so, it is important to gauge how accurate the event reconstruction algorithm is to reality, especially considering that complicated secondary effects such as light scattering intensity in water and photomultiplier tube responses may not be well-modeled in the detector simulation.

\section{\label{sec:skdetsyst}SK Detector Systematic Errors}
\subsection{\label{subsec:systerrparams}Systematic Error Parametrization in SK}

\begin{figure*}[b!]
    \centering
    \begin{subfigure}[b]{0.3\textwidth}
        \centering
        \includegraphics[height=40mm]{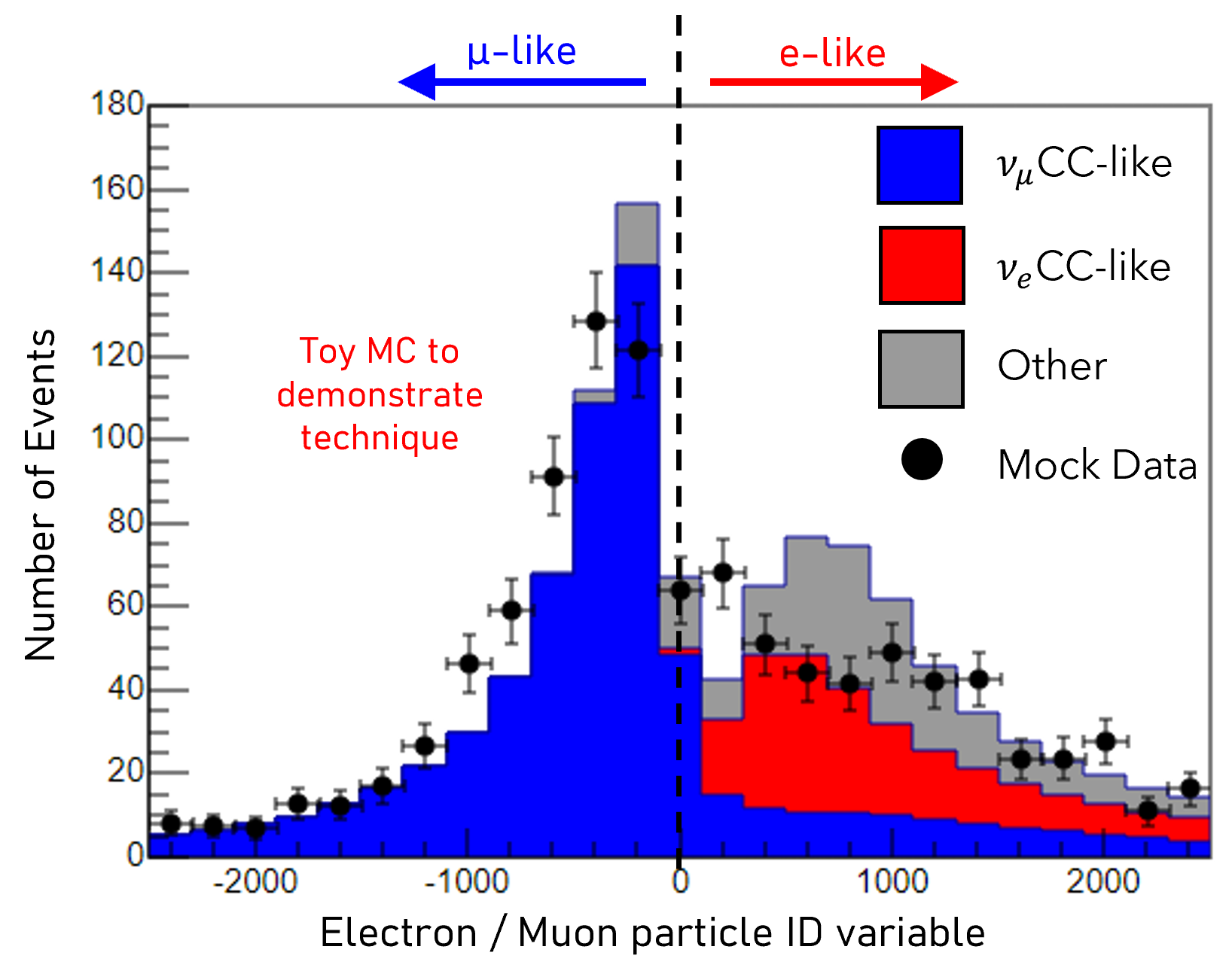}
        \caption{$\alpha=1,\ \beta=0$}
        \label{subfig:emu_base}
    \end{subfigure}
    ~
    \begin{subfigure}[b]{0.3\textwidth}
        \centering
        \includegraphics[height=40mm]{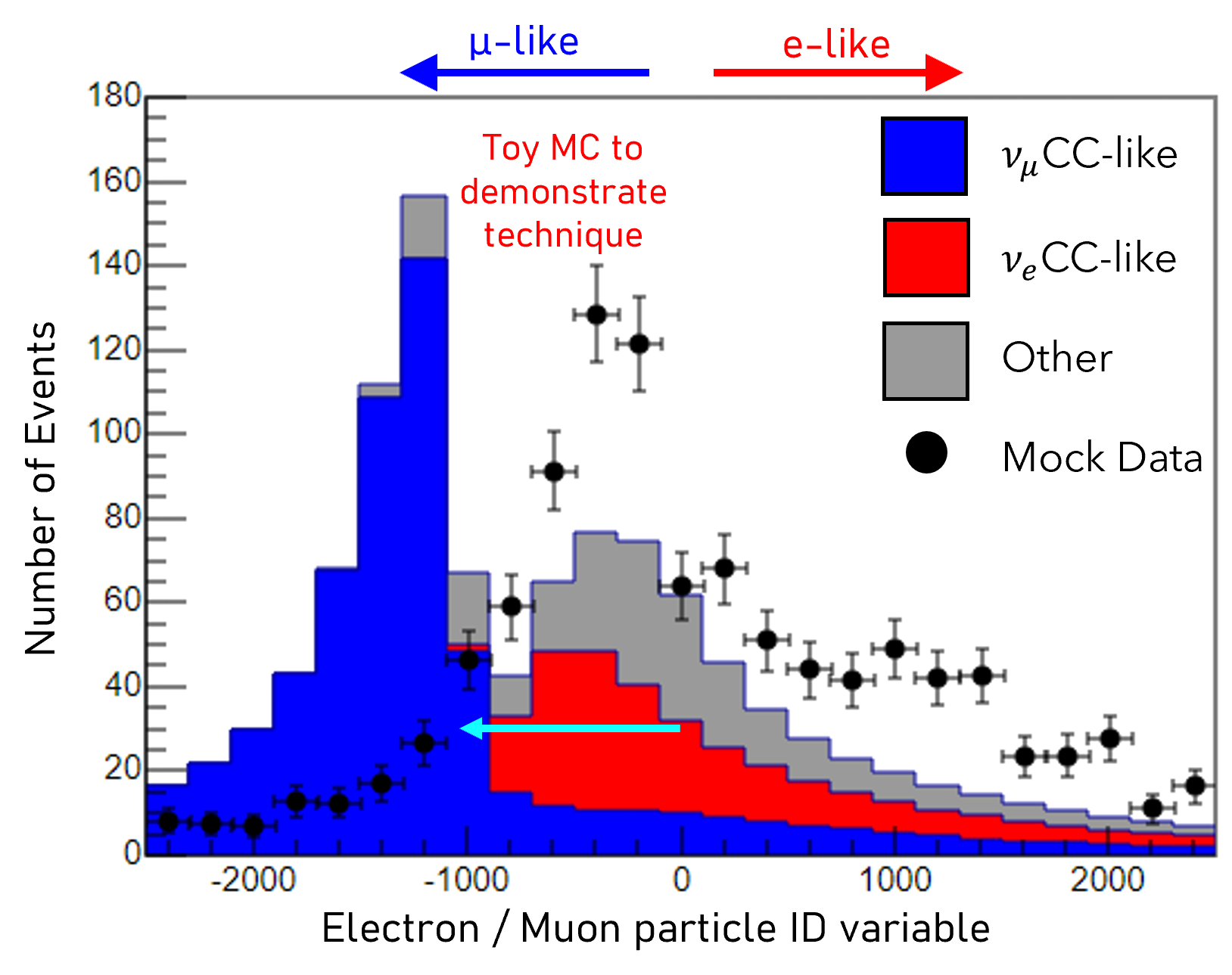}
        \caption{$\alpha=1,\ \beta<0$}
        \label{subfig:emu_neg_beta}
    \end{subfigure}
    ~
    \begin{subfigure}[b]{0.3\textwidth}
        \centering
        \includegraphics[height=40mm]{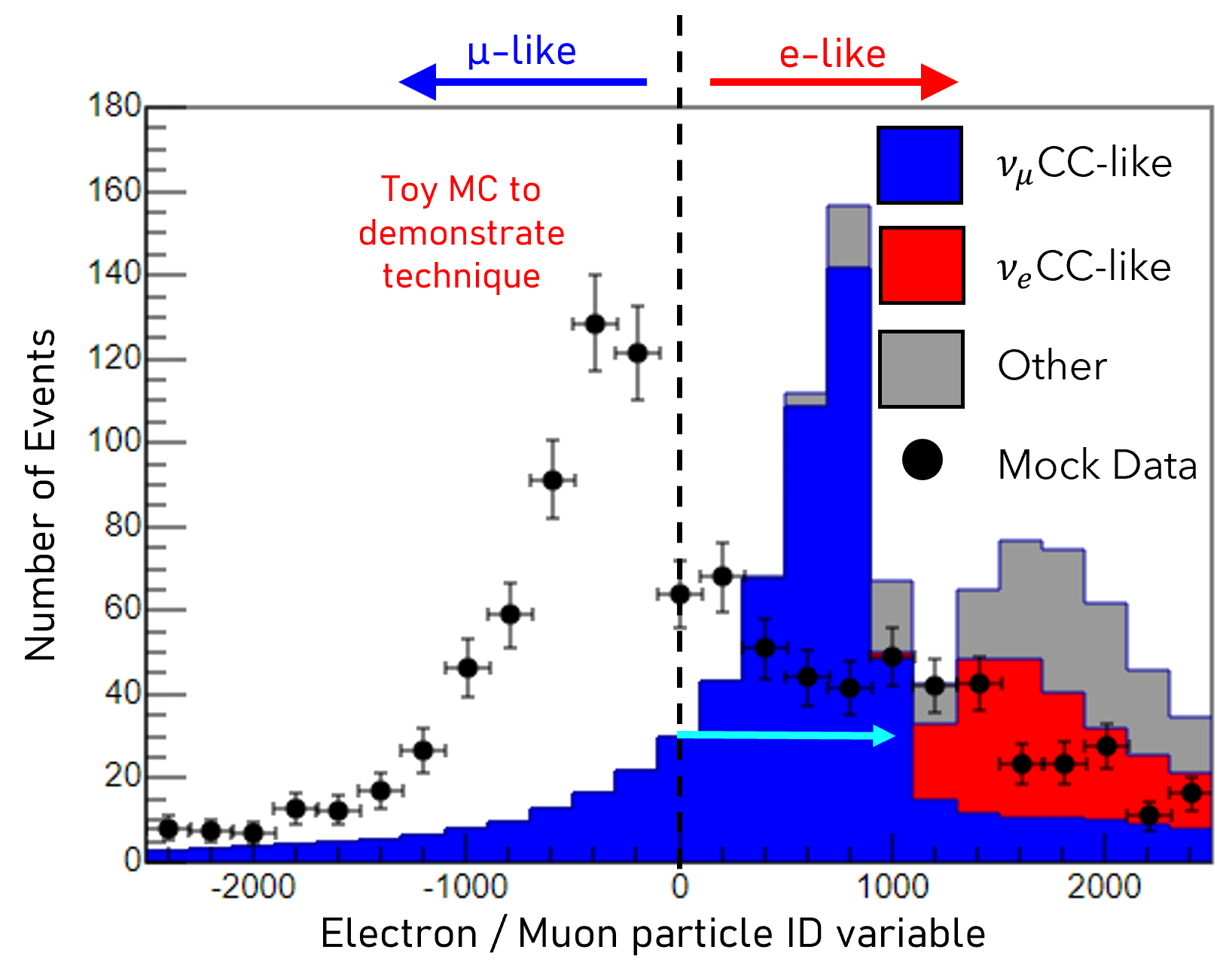}
        \caption{$\alpha=1,\ \beta>0$}
        \label{subfig:emu_pos_beta}
    \end{subfigure}
    ~
    \begin{subfigure}[b]{0.3\textwidth}
        \centering
        \includegraphics[height=40mm]{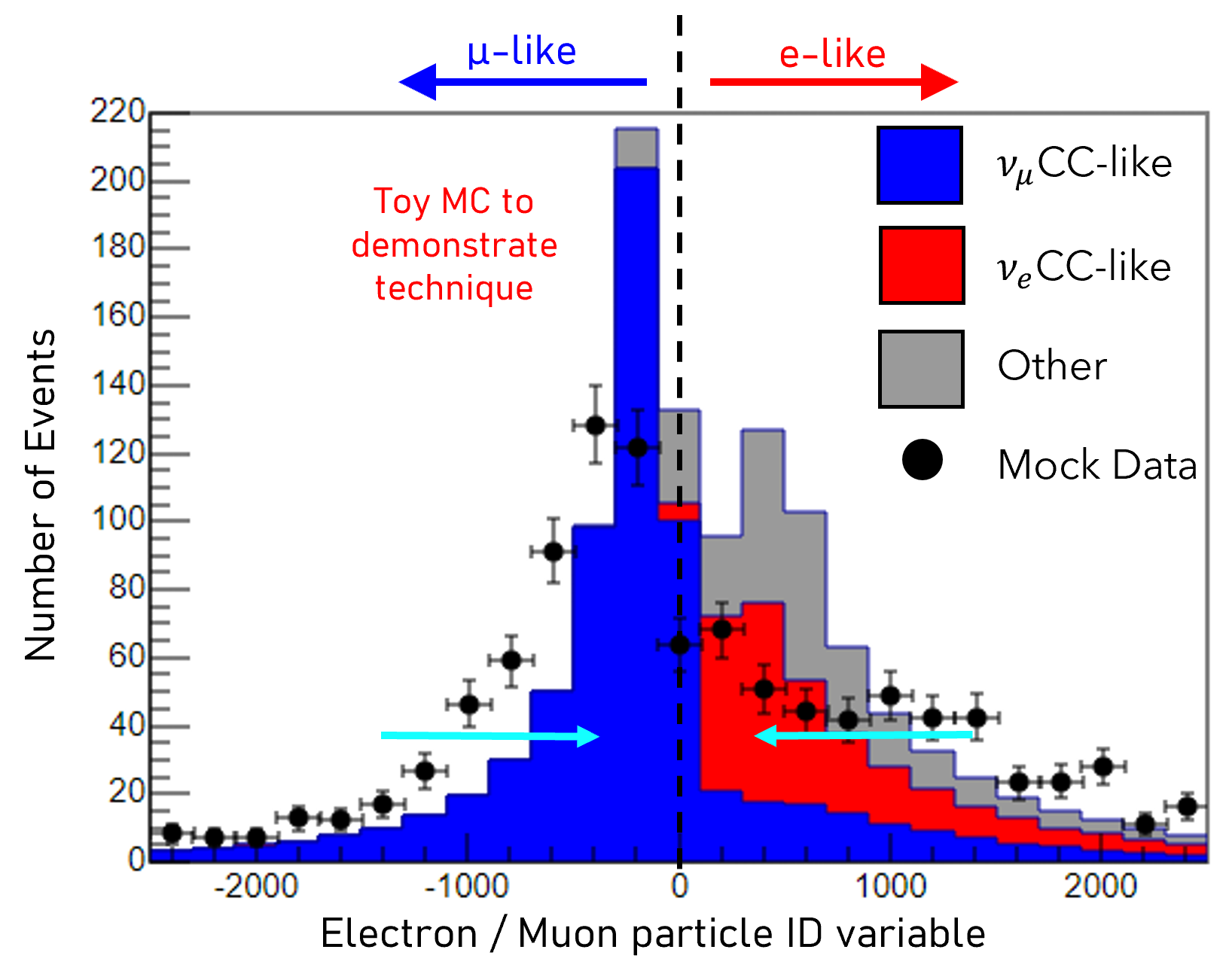}
        \caption{$\alpha<1,\ \beta=0$}
        \label{subfig:emu_neg_alpha}
    \end{subfigure}
    ~
    \begin{subfigure}[b]{0.3\textwidth}
        \centering
        \includegraphics[height=40mm]{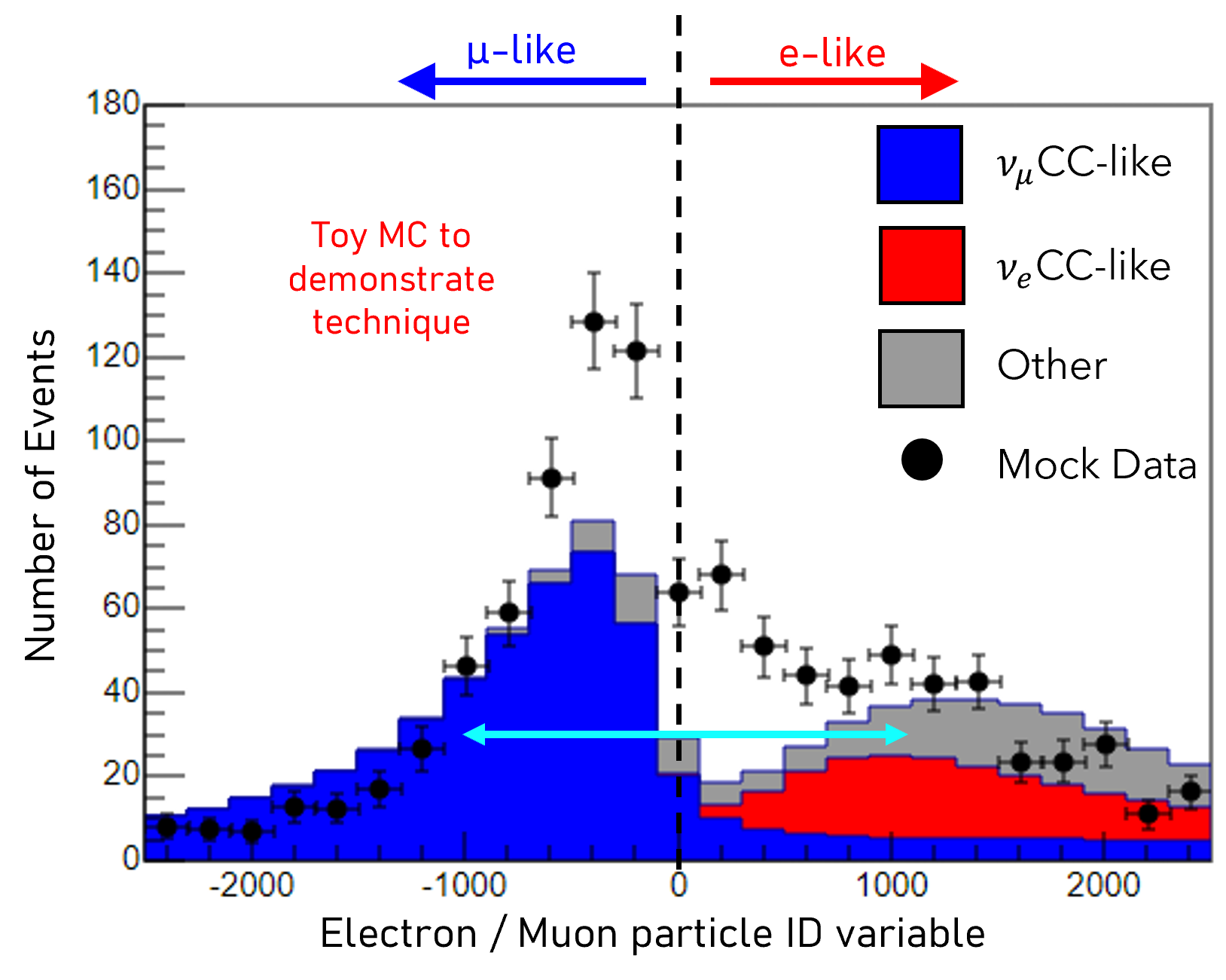}
        \caption{$\alpha>1,\ \beta=0$}
        \label{subfig:emu_pos_alpha}
    \end{subfigure}
    \caption{Toy data vs MC histogram comparisons binned in terms of the $e/\mu$ PID variable $L$. The PID values of the MC events are modified using $L\rightarrow\alpha\cdot L+\beta$ to demonstrate the shifting and smearing technique.}
    \label{fig:example_PID}
\end{figure*}

The T2K far detector event selection is performed using log-likelihood ratio-based particle identification (PID) variables. Of interest here are the PID variables that are used to select signal samples for the T2K oscillation analysis; for example, there is an $e/\mu$ PID variable for distinguishing between the $e-$like and $\mu-$like analysis samples. T2K compares reconstructed data to Monte Carlo (MC) simulation using histograms binned in the PID variables of interest, as demonstrated in Figure~\ref{subfig:emu_base}, to get a measure of the relative uncertainty in reconstruction. Then to parametrize this uncertainty, T2K assumes that any underlying mis-modeling in SK will manifest at the analysis level as systematic shifts in the PID values for data events relative to simulated ones. Consequently, T2K uses a pair of catch-all variables to recover this behavior -- a smearing parameter $\alpha$ and a shifting parameter $\beta$ -- that modify the nominal PID value $L$ of simulated events like so:
\begin{align}
    \label{eqn:alpha_beta_modification}
    L \rightarrow \alpha\cdot L + \beta.
\end{align}
By using this method to modify the MC distributions to better match the data, this parametrization effectively encodes the systematic shifts experienced by the reconstructed data into our $\{\alpha,\beta\}$. Additionally, using such a general method allows us to accommodate a wide variety of effects while remaining agnostic to the specific source of underlying uncertainty. This makes the technique robust against changes to the detector and improper modeling choices. Examples of how these ``shifting'' and ``smearing'' variables can affect the simulated distributions are shown in Figures~\ref{subfig:emu_neg_beta}--\ref{subfig:emu_pos_alpha}.

\subsection{\label{subsec:atmfit}Atmospheric Neutrino Fit}
To estimate the SK detector mis-modeling between data and simulation, T2K utilizes atmospheric neutrino data and MC as its ``control'' samples. SK has a robust atmospheric neutrino program with nearly 30 years worth of data, which makes for a strong control data set that spans the energies and event types that are relevant to T2K. It is assumed that any mis-modelings observed between the SK atmospheric neutrino data and MC will be applicable to the T2K beam neutrino model due to the overlap in kinematic range, event topology, and detector condition between the beam and atmospheric regimes. Thus, systematic uncertainties are calculated using SK atmospherics and then transferred to the T2K beam analysis.

Mechanically, this so-called atmospheric neutrino fit is performed using a Markov Chain Monte Carlo (MCMC) framework. The MCMC proposes different values of $\{\alpha,\beta\}$, which modify the PID values of the MC on an event-by-event basis according to Equation~\ref{eqn:alpha_beta_modification}. The modified MC is then binned into histograms in terms of the various PID variables as described in Section~\ref{subsec:systerrparams}. A shape likelihood is calculated between the data and modified MC histograms, acting as the optimization function for the MCMC. By repeating this for many $\{\alpha,\beta\}$, this process effectively encodes the differences between the data and MC into a posterior set of $\{\alpha,\beta\}$.

\begin{figure*}[b!]
    \centering
    \begin{subfigure}[b]{0.45\textwidth}
        \centering
        \includegraphics[height=50mm]{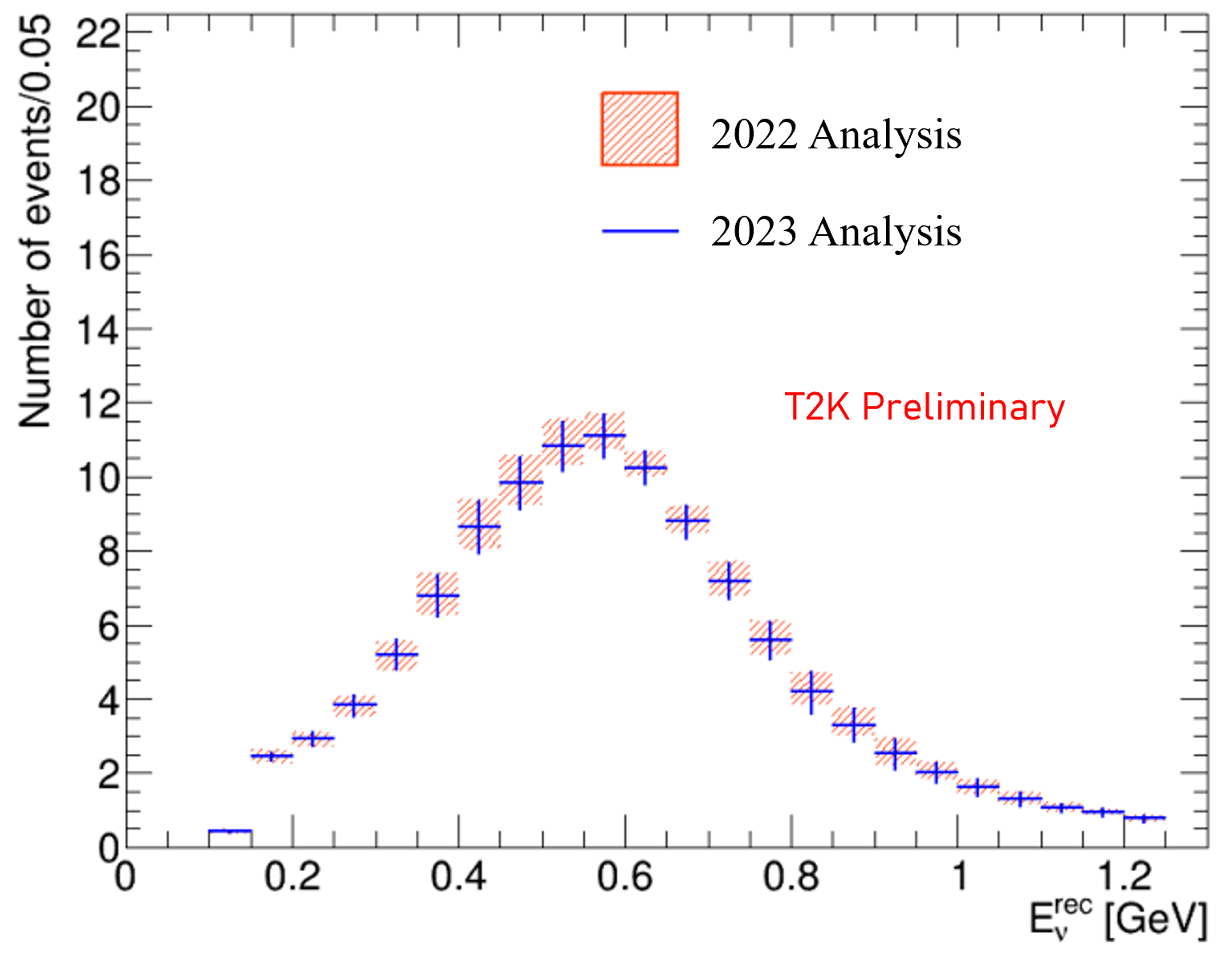}
        \caption{Single ring $e-$like}
        \label{subfig:sr_elike_post_predictive}
    \end{subfigure}
    ~
    \begin{subfigure}[b]{0.45\textwidth}
        \centering
        \includegraphics[height=50mm]{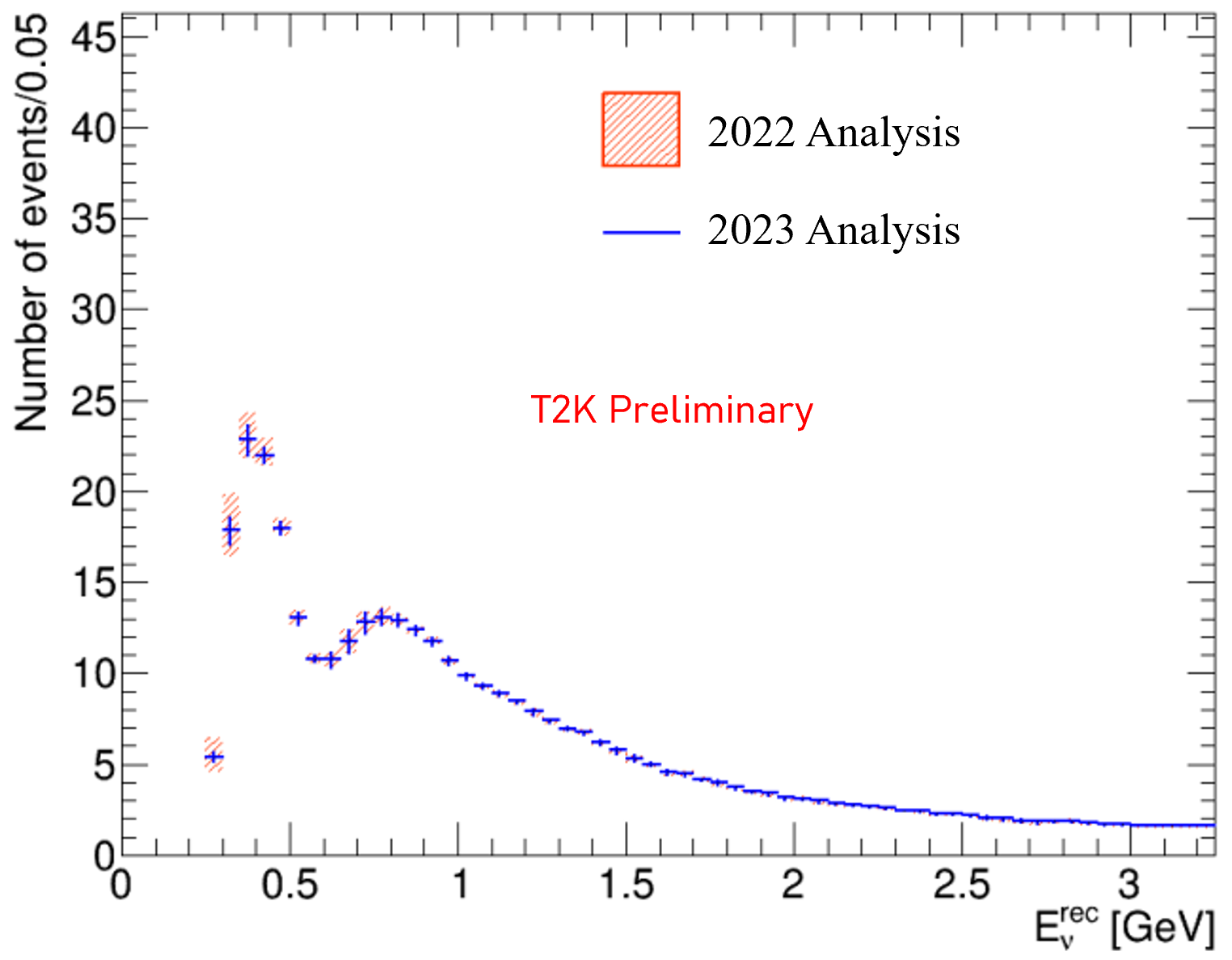}
        \caption{Single ring $\mu-$like}
        \label{subfig:sr_mulike_post_predictive}
    \end{subfigure}
    \caption{Posterior predictive energy spectra for different T2K analysis samples. Bin errors denote the 1$\sigma$ uncertainty coming from the SK detector error matrix. Differences in bin error here are largely attributed to the removal of several ``ad-hoc'' systematics between the 2022 and 2023 analyses}
    \label{fig:posterior_predictives}
\end{figure*}

\subsection{\label{subsec:errormatrix}SK Detector Error Matrix}
To make the output from the atmospheric neutrino fit described in Section~\ref{subsec:atmfit} usable to the T2K oscillation analysis, the $\{\alpha,\beta\}$ from the atmospheric neutrino fit are applied to the T2K beam MC in a toy MC procedure. In this toy MC procedure, random sets of $\{\alpha,\beta\}$ are taken from the MCMC posterior and used to modify the PID values of beam MC events. Because events are selected into T2K samples based on the values of their PID variables, the modified events will migrate in and out of the T2K analysis samples. This will change the observed neutrino energy spectrum, as demonstrated in Figure~\ref{fig:posterior_predictives}. This uncertainty in the observed posterior spectra is encoded into a covariance matrix, binned in terms of the different T2K samples and neutrino energies. The so-called SK detector error matrix is then used in the T2K oscillation analysis to provide an uncertainty in the observed beam neutrino energy spectrum at SK.

\section{\label{sec:outlook}Outlook}
The SK detector systematic analysis has made several improvements that will impact future T2K analyses. First, T2K is planning to add a new $\nu_eCC1\pi^+$ sample to increase $\nu_e$ signal statistics at SK, and a new $NC\pi^0$ sample to help constrain $\pi^0$ backgrounds in the $\nu_e$ samples. Infrastructure for estimating detector systematics for these new samples has been added to the detector systematics framework. Additionally, the detector systematics analysis has moved to a new MC production for the first time in 5 years, complete with a new neutrino interaction model described in~\cite{oa2021paper}. Finally, improved implementation techniques have drastically improved the convergence time of the analysis from O(months) to O(weeks). In the absence of new data, the treatment of these SK systematic uncertainties will be among the major updates for the next T2K oscillation analysis.


\bibliography{NuFact_Proceedings}

\end{document}